# Privacy of the Internet of Things: A Systematic Literature Review (Extended Discussion)

Noura Aleisa and Karen Renaud
School of Computing Science, University of Glasgow
n.aleisa.1@research.gla.ac.uk, karen.renaud@glasgow.ac.uk

*Abstract*—The Internet of Things' potential for major privacy invasion is a concern. This paper reports on a systematic literature review of privacy-preserving solutions appearing in the research literature and in the media. We analysed proposed solutions in terms of the techniques they deployed and the extent to which they satisfied core privacy principles. We found that very few solutions satisfied all core privacy principles. We also identified a number of key knowledge gaps in the course of the analysis. In particular, we found that most solution providers assumed that end users would be willing to expend effort to preserve their privacy; that they would be motivated to act to preserve their privacy. The validity of this assumption needs to be proved, since it cannot simply be assumed that people would necessarily be willing to engage with these solutions. We suggest this as a topic for future research.

## I. INTRODUCTION

With the growth of the Internet of Things (IoT) your future morning routine might be something similar to the following scenario:

*It is morning; your smart home is readying itself to support your daily routine. The alarm finds out when you have to get up by accessing your diary, it knows how long it usually takes you to get out of the house, based on the data collected from your phone, fine-tuned by consulting timings from previous days. The light is switched on, and the coffee machine starts brewing your daily dark roast. You wake, dress and eat breakfast. Your autonomous car has started itself, reversed out of the garage, and is waiting for you to hop in. On your way out, your Smartphone locks the door and activates the alarm. Your refrigerator adds 'milk' to your convenience store shopping list, so that your parcels will be ready for you to pick up on your way home from work.*

*During your journey to work your autonomous car drives itself, using millions of embedded sensors. It goes directly to the parking spot it has detected using a networked application that receives notifications from the city's parking bay sensors.*

This, then, is the wonderful new world of the Internet of things [28], [12].

The term "Internet of Things" was first used by Kevin Ashton at Procter & Gamble in 1999, to describe an Internet-based information service architecture [3]. Generally the term refers to Internet-enabled objects interacting with each other and cooperating to achieve specific goals. These objects could be RFID, sensors, actuators or mobile phones [21]. The Internet of Things claims to improve peoples' lives. For instance, a tool could measure heart rate and body temperature, and then communicate with the energy management system to adjust room temperature depending on the individual's physiological status [62]. Other tools activate smart streetlights, monitor surveillance cameras and control traffic lights. Collected information can be shared with different stakeholders to improve business intelligence [75].

The IoT makes life less effortful and more convenient. On the other hand, the invisibility of the data collection, usage and sharing processes raise concerns. The privacy of IoT users could easily be sacrificed [17]. On the one hand, we accept the fact that the service providers need to access our information in order to deliver tailored services. On the other hand, we also expect our private information to be protected from unauthorized access, and not shared with 3rd parties [64].

The contribution of this paper is to provide an overview of existing IoT privacy-related research in order to identify areas of focus and highlight areas that deserve more attention.

## II. PRIVACY

### A. Definition

Solove has defined privacy as "*an umbrella term, referring to a wide and disparate group of related things*" [61] (p.485). Privacy, according to Privacy International, is a multidimensional concept, which is related to four components: (1) body, (2) communications, (3) territory, and (4) information. Bodily privacy focuses on the people's physical protection against any external harm. Privacy of communications focuses on the protection of the information that is carried through any medium between two parties. This includes email, mail and telephone. Territorial privacy is about establishing boundaries or limits on physical space or property, such as the home, workplace, and public places. Information privacy refers to personal data that is collected and processed by an organization, such as medical records and credit card information [63].

### B. Privacy Stances

Westin's take on privacy is that of someone having the right to control what personal information collected about them or known to others [76]. As technology makes it trivial for organizations to maintain comprehensive digital files about every person, privacy concerns have emerged. People are concerned about what data is collected, who has access to it, who controls it, and what it is used for [37]. Westin carried out studies to study privacy perceptions between 1978 and 2004 and created a "Privacy Index". Westin said that people naturally fell into one of three categories with respect to their privacy stance: *Fundamentalist, Pragmatist* and *Unconcerned* [35]. Fundamentalists are concerned about the accuracy of collected information and uses made of it. They are generally in favour of laws supporting privacy rights as well as enforceable privacy-protecting frameworks. Pragmatists are willing to give some personal information to a trusted service provider in return for benefits. Unconcerned people have full trust that the organizations collecting their information would not abuse it.

Westin's follow-up surveys revealed that the percentage of "Unconcerned" had decreased over the last few years. He attributes this to people becoming more aware of technology and different means of preserving their privacy. It could also indicate an increasing level of concern about privacy [35]. A number of privacy breaches have made headlines in recent years. For example, this year it was reported that unsecured webcams exposed the private lives of hundreds of consumers on the Internet [52]. Hewlett Packard's 2015 report [27] reported that 80% of IoT devices raised privacy concerns.

### C. Privacy Threats

Nowadays, it is even harder for us to retain our privacy, as the Internet of Things technologies take over our daily lives. Conflicts over how organizations can access individual data are pervasive, and IoT will add to this. Ziegeldorf's literature review [84] enumerates the most common privacy threats in the Internet of Things:

1) *Identification* is the most dominant threat that connects an identifier, e.g. a name and address, with an individual entity;
2) *Localization and tracking* are the threat of locating an individual's location through different means, e.g. GPS, internet traffic, or smartphone location;
3) *Profiling* is mostly used for personalization in e-commerce (e.g. in newsletters and advertisements). Organizations compile information about individuals to infer interests by association with other profiles and data sources;
4) *Interaction and presentation* refers to the number of smart things and new ways of interacting with systems and presenting feedback to users. This becomes a threat to privacy when private data is exchanged between the system and the users;
5) *Lifecycle transitions* occur when an IoT items is sold, used by its owner and finally disposed of. There could be an assumption that all information is deleted by the object, but smart devices often store huge amounts of data about their own history throughout their entire lifecycle. This could include personal photos and videos and are sometimes not deleted upon transfer of ownership;
6) *Inventory attacks* apply to the unauthorized access and collection of data about the presence and characteristics of personal things. Burglars can use inventory data to case the property to find a safe time to break in;
7) *Linkage* consists in linking different systems, the chance of unauthorized access and leaks of private data grow when systems are linking to combine separate data sources.

### D. Privacy Principles

The ISO [32] and the OECD [46] have identified 11 privacy principles from privacy laws and regulations based on the international guidelines that have been defined to protect privacy. Wright and Raab [78] extend that to 20 principles. They argue that these principles be considered as new products and services are developed.

Some of the principles are particularly applicable to IoT, such as "Right to confidentiality and secrecy of communications" (violated by Samsung [42]), "Consent and choice" (violated by LG [60]) and "People should not ... be denied goods or services or offered them on a less preferential basis" (violated by Toshiba [50]). It seems as if the IoT developers have not taken Wright and Raab's [78] admonition to heart, hence the need for privacy-related IoT privacy-preserving solutions.

### E. Privacy Preserving Solutions

In order to address the privacy concerns of end-users and privacy considerations of service providers, several approaches have been proposed by the research community:

1) *Cryptographic techniques and information manipulation*: Although researchers have spent many years proposing novel privacy-preserving schemes, cryptography is still the dominant one in most current proposed solutions, even though, for all of the obstacles they may face, many of the sensors cannot offer adequate security protocols due to the limited amount of storage and computation resources [16].
2) *Privacy awareness or context awareness*: Solutions for privacy awareness have been mainly focused on individual applications that provide a basic privacy awareness to their users that smart devices, such as smart TVs, wearable fitness devices, and health monitor systems could collect personal data about them. For instance, in recent research, a framework called SeCoMan was proposed to act as a trusted third party for the users as applications might not be reliable enough with the location information that they manage [31].
3) *Access control*: Access control is one of the viable solutions to be used in addition to encryption and privacy awareness. This gives users the power to manage their own data. An example of this approach is CapBAC [59], proposed by Skarmeta, Hernandez, and Moreno. It is essentially a distributed approach in which smart things themselves are able to make fine-grained authorization decisions.
4) *Data minimization*: The principle of "data minimization" means that the IoT service providers should limit the collection of personal information to what is directly relevant. They should also retain the data only for as long as is necessary to fulfill the purpose of the services provided by the technology. In other words, they should collect only the personal data they really need, and should keep it only for as long as they need it [66].

There are other proposed solutions that do not fall into the previous four categories, such as *hitchhiking*. This is a new approach to ensure the anonymity of users who provide their locations. Hitchhiking applications handle locations as the entity of interest. Because the knowledge of who is at a particular location is unnecessary, the fidelity tradeoff is removed [68].

Another example is the *introspection* technique that proactively protects users' personal information by examining the activities of the VM. It gathers and analyzes the CPU state of every VM, the memory contents, file I/O activity, network information that is delivered via hypervisor and detects malicious software on the VM. However, if IoT device loses integrity due to any malicious attack, it creates risks to the users' privacy [34].

## III. METHODOLOGY

To assess the limits of privacy that are potentially violated by the Internet of Things, a systematic quantitative literature review was conducted. This method, according to Pickering and Byrne [49], has benefits as compared to a narrative style. It is capable of identifying the areas covered by existing research, and also revealing the gaps. It approaches the literature from different perspectives and facilitates delivery of new insights. Figure 1 depicts the process.

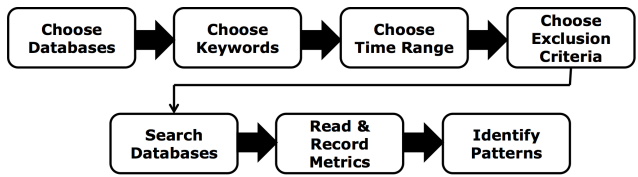

Fig. 1. Systematic Literature Review Process

*Choose Databases:* Papers published in academic journals were collected from electronic databases, including Google Scholar, Web of Science, ProQuest, Research Gate, SCOPUS, and Science Direct.

*Choose Keywords:* Keywords used for the searches were 'Internet of Things', 'IoT', and a combination of terms including: 'privacy', 'trust', 'awareness', 'data', 'protection', 'security', 'preserving', 'individual', 'user', and 'private'.

*Choose Time Range:* The search was restricted to papers, published between 2009 and 2016.

*Choose Exclusion Criteria:* The academic search was restricted to papers published in English. In addition to the research papers, a search for news stories and privacy reports were also included in order to accommodate personal privacy violation perspectives. Review papers were excluded but their reference lists were followed to ensure all the research in this field was consulted.

*Searching & Recording:* For each collected paper, the following information was recorded including author(s), year of publication, journal, country where the research was carried out. Each paper was categorized based on the methods used and whether analysis was quantitative, qualitative, or mixed. The rest of the criteria are related to the researched topic, it classifies the application area as home automation, smart cities, smart manufacturing, health care, automotive, or wearable devices, the type of technology used (RFID, sensor, nano, or intelligent embedded technology). The privacy protections, threats, violations, and perceptions for each type of technology were also recorded. Perceptions were categorized based on Westin's three categorizes: fundamentalist, pragmatic, and unconcerned [35].

*Identifying Patterns:* An analysis was carried out to uncover patterns in order to identify foci, gaps and to make recommendations for future research.

## IV. RESULTS

A total of 122 original research papers on the privacy of the Internet of Things were identified (Table III in the Appendix). In this section, the geographic scope, characteristics and methods, threats, solutions, and user privacy perceptions are presented.

### A. Geographic scope

Privacy research was carried out by 26 countries with Europe dominating: most papers were from Germany (19.6%), Italy and France (12.5%).

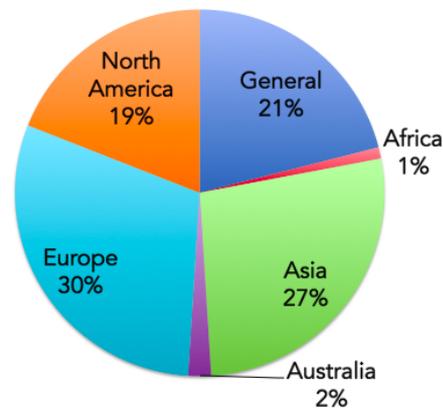

Fig. 2. Paper Locations

### B. Methods used by Researchers

A wide range of methods have been used to assess the privacy of the IoT. Many studies used multiple methods to collect data. Based on the methods sections in Table III, almost 52 (44.1%) papers used modeling, while only 16.9% of studies used document analysis, followed by case studies (15.2%), surveys (12.7%), observation (10.1%), and interviews (0.8%). Nearly half of the studies (45.4%) adopted quantitative research strategies, with a few using a qualitative approach (19.8%), and mixed approaches (16.5%). Another type of data has been considered here, with 18.2% for news or reports.

## C. Characteristics of IoT

Papers often assessed the characteristics of the Internet of Things, including: technologies used in the IoT, application areas, and types of privacy protection. When papers specified what technologies were used in the IoT, most discussed the use of RFID (34.9%) and sensor technology (55.3%). Further consideration shows that 37% were about home automation, then smart cities (16.8%), and the remainder fluctuated between 13.6% and 9.6% for automotive, health care, wearables, and manufacturing (Table III in the Appendix).

One of the key concerns for users are concerns about the secure services offered by the IoT technology. The review has provided a comparison between security and privacy protection solutions and the individual's perceptions of the IoT. In terms of the level of security protection, most papers (66.6%) have mentioned that the authentication and authorization techniques are the most common security practices used in the IoT. On the other hand, the review has found that there was an increase in three privacy protection mechanisms, with 39.5% for cryptographic techniques and information manipulation, 26.1% for privacy awareness or context awareness, and 25.5% for using access control (Table III).

Most of the reviewed research considers the lack of privacy protection a major challenge. 48% of the solutions were for home automation smart products, then for health care (20%), then for automotive, smart cities (12%), and the remaining 4% for wearables and manufacturing.

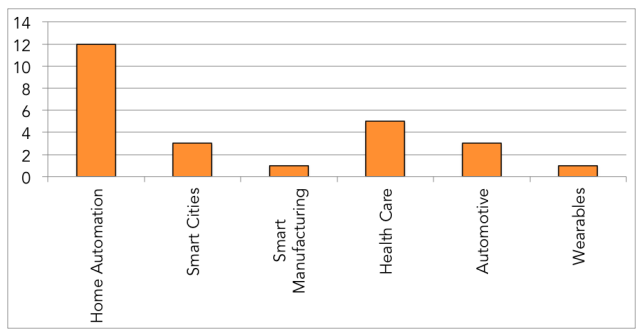

Fig. 3. IoT Privacy Protection Solutions

## D. Threats, Solutions, Principles, Perceptions

The increasing collection of data about individuals is one of the main concerns identified in most of the papers, especially the threats to individuals caused by analysis of their data using data mining techniques [10]. The literature indicates that about 31.5% of the papers have concerns about location tracking; the next concern for individuals is the sharing of unanonymised data (25.9%). Concerns about profiling have been mentioned in 21.3% of the papers, followed by inventory attacks (8.3%), interaction and presentation (6.5%), life cycle transitions (3.7%), and linkage (2.7%) (Figure 4) [84].

A wide range of approaches have been proposed to conserve user privacy in IoT. Over half of these have not been tested or evaluated; they are essentially at the proposal stage. On the other hand, about 39 solutions were evaluated namely: cryptographic algorithms, control access management tools, data minimization techniques, and privacy or context awareness protocols (Table I).

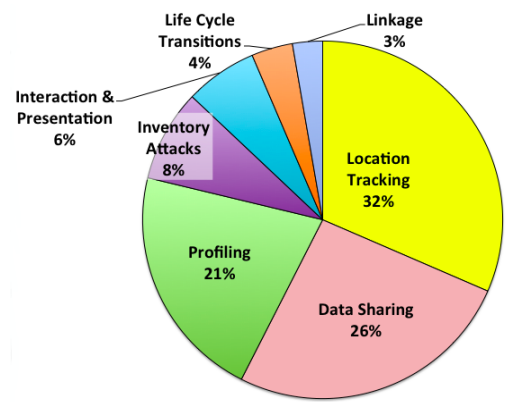

Fig. 4. Highlighted IoT Privacy Threats

TABLE I
SUMMARY OF LITERATURE: FIVE KEY THEMES OF SOLUTIONS HIGHLIGHTED BY THE IoT REGARDING INDIVIDUAL PRIVACY

| Privacy Protection Themes | Total | Literature Reference |
|---|---|---|
| **Tested and Evaluated** | | |
| Cryptographic techniques and information manipulation | 15 | [1], [67], [44], [36], [18], [53], [22], [5], [64], [30], [7], [13], [77], [59], [73] |
| Data minimization | 3 | [15], [7], [59] |
| Access control | 6 | [1], [31], [30], [13], [39], [59] |
| Privacy awareness or context awareness | 12 | [1], [31], [55], [5], [81], [7], [71], [70], [77], [69], [59], [8] |
| Differential Privacy | 0 | |
| Other (introspection, trust assessment and evaluation) | 3 | [34], [6], [40] |
| **Not Evaluated** | | |
| Cryptographic techniques and information manipulation | 16 | [24], [25], [79], [9], [20], [51], [41], [54], [43], [26], [45], [58], [47], [19], [2], [56] |
| Data minimization | 2 | [25], [19] |
| Access control | 14 | [25], [9], [57], [20], [41], [54], [26], [45], [58], [72], [47], [19], [29], [74] |
| Privacy awareness or context awareness | 9 | [25], [41], [33], [54], [72], [4], [2], [56], [11] |
| Differential Privacy | 1 | [19] |
| Other (multi-routing and random walk, hitchhiking) | 2 | [83], [68] |

(Table II in the Appendix) shows the 11 privacy principles that have been identified by OECD [46], and it has determined for each of the proposed solutions to protect individuals' privacy in IoT the principles that have been focused on. As It can be noticed that only 4 out of 75 of the solutions have considered all the privacy principles on their proposed model, and only 9 have focused on 10 principles. The rest of the solutions have focused on about only 6 of the privacy principles.

Under the assumptions that individuals do not have the power to control their own data and protect their own privacy, the privacy violation of the data collected by smart devices has become a major concern to the public. To represent these concerns, we classified and recorded the collected papers according to the respondents using the Westin's categories as follows.

We counted most of the papers that offered privacy-preserving frameworks, discussed the privacy threats, or even demonstrated concern about the data collected and used by IoT as fundamentalist. With regard to pragmatic, we allocated papers that encourage trusting the privacy and security level of the smart devices, without having any awareness about the personal collected data, to this category. Two papers argued for the benefits of a smart environment and used the "nothing to hide" argument — these were unconcerned authors.

The majority of the research papers can be classified as fundamentalist (112 out of 122 papers, including news and reports that were written by non-specialists), while only 6 papers are pragmatic, and 2 papers demonstrate unconcern. Having the majority of the papers under the first category can be explained due to that most of the papers were written by privacy professionals proposing models to protect individuals privacy in IoT.

## V. Discussion

The literature has presented insights into where, how and what research has been conducted and made it possible to identify the gaps.

### A. Primary Research Focus

As shown in Fig. 2, most research, to date, has been conducted in Europe, and Asia, within non-English speaking countries such as Germany, Italy, Spain, France, India, and China. This shows that the individual privacy concerns are not limited to English-speaking countries.

The results suggest that countries with the strictest personal privacy measures, such as those in Europe, seem to do the most research in this area [23].

The deployed study methods fell into one of two categories: (1) analyzing the privacy violations and threats, and (2) proposing a solution to protect the IoT user's privacy. Modeling, document analysis and case studies are dominant. In contrast, few observational or survey-type studies were carried out on privacy breaches and perceptions.

The range of research demonstrates a growing awareness of the potential for privacy violation. Researchers have started exploring privacy protection mechanisms. The sheer range and variety of IoT products, each on bespoke platforms, makes this a challenging field to find solutions for.

Most of the papers examined in this systematic review were published in academic venues. However, a number of news reports were also included to gauge consumer concerns about privacy as well. It can reasonably be concluded that such concerns are not only being raised by technology professionals but by consumers with less technological expertise.

### B. Threat Focus

The majority of the reported threats were focused on data being collected about individuals themselves, such as their identities, location, or profiling. This information can be used to harm the users, to carry out identity theft, or burglaries.

Figure 3 shows that the majority of proposed privacy-protecting applications and techniques are for smart devices used in homes or for health monitoring. These include Smart TVs, Smart Meters, light or temperature control, Smart remote health monitors, or drug tracking. Such a restricted focus could be attributed to several circumstances including: (1) the availability and the easy access of the homes or health care smart devices in the market; (2) The homes or health care smart devices are not required to be controlled by higher authority as in the smart cities and manufacturing which controlled by government or private organizations; (3) growth of automotive, cities, and (4) manufacturing smart technology has not become reality yet.

### C. Gaps

Many of the new applications or techniques proposed to protect individual privacy will intimately involve humans in the process. Some solutions deploy access control methods, or privacy-awareness applications. For example, in [71], the study proposed the Dynamic Privacy Analyzer (DPA), a solution to make the smart-meter data owner aware of the privacy risks of sharing smart meter data with third parties. On the other hand, almost half of the proposed solutions proposed taking the human out of the loop. These proposed using cryptographic techniques and information manipulation, or data minimization to prevent data being sniffed *en route* to servers. In [67], an original scheme called the Path Extension Method (PEM) was presented, which provides powerful protection of source-location privacy, by using an encryption technique that ensures an adversary will not be able to eavesdrop on communications.

The overwhelming majority of the researchers were fundamentalist about privacy. This is, perhaps, to be expected since unconcerned researchers would not have any interest in carrying out research in this area. It does mean, however, that they might be somewhat unrealistic about the man and woman in the street, and their privacy stance. Unconcerned consumers are likely to be unwilling to take any actions at all to preserve a privacy they don't care about. Solutions seem to be designed under the assumption that consumers will naturally be willing to spend time and effort engaging with them. This assumption might well be flawed.

The question that demands investigation is whether consumers of various privacy stances will be willing to expend effort to interact with privacy-preserving applications. Researchers are coming up with innovative solutions but this will be futile in the face of consumer complacency or unwillingness to engage with them.

### D. Returning to Privacy Principles

Table II shows how the different solutions map to the privacy principles. It can be observed that only a few solutions cover all 11 principles; the average coverage is 6 principles. The two principles that almost all the solutions deliver are security and integrity/accuracy. While protection from unauthorized access, modification of data, and ensuring accuracy are very important, this does not make the other principles less important. One of the least-considered principles is the *Purpose specification*. Designers do not seem to believe this is one of the user's rights, i.e. knowing why the smart device needs the particular data they are collecting.

The results demonstrate that designers' priorities are to secure the collected data, to ensure that it is accurate and updated, and not transferred without protection. It is time for them to pay more attention to designing for privacy awareness and enabling protection thereof.

Privacy is all about the user; most of the principles mandate his/her involvement, entailing notification of the device policy, the data collected, the purpose of collecting specific types of information, giving him/her the ability to control information disclosure. He/she can also ensure that the data is not going to be used for purposes other than that specified in the policy, and that collection of personal information is minimised. Having the user involved from the outset is the best way to gain trust.

*E. Need for Legislation*

A significant number of ambiguities remain poorly described in the literature, and require further investigation. For example, consumers would sometimes like to know what data is recorded and transmitted by their smart device before they buy it. This is not currently possible. It would also be helpful if the consumer could get information about how their data is protected by the device, both on the device itself, and during transmission. This information is not generally provided. Finally, devices ought to allow people to configure privacy preferences, in much the same way as Smartphones and Facebook currently allow people to, but perhaps because of the newness of this technology, this functionality is not offered. It is clear that the industry is going to have to be compelled to respect privacy. Their track record so far amply demonstrates that they do not have the will to do this without some motivation to do so.

## VI. LIMITATIONS

Although the Smartphone qualifies as an IoT device it was not explicitly included in the search keywords. We wanted to focus on papers that claimed to solve IoT-wide issues, not those focusing only on one type of device.

This review has focused primarily on privacy-related research. In some cases it is difficult to separate privacy- and security-preserving solutions. For example, encryption is primarily a security tool, but, if used, essentially preserves the privacy of communication. A further review should be carried out in order to analyze security-specific IoT solutions as well.

## VII. RELATED RESEARCH

The Internet of Things is considered a significantly disruptive technology of this era, because it integrates several collaborative technologies, allowing for comprehensive data collection. The IoT allows third parties to collect and analyse data about the environment and individuals traits, allowing the delivery of personalised services that require no deliberate interaction [10].

Opplinger [48] refers to the difficulties of preserving security and privacy because the IoT has no boundaries. In introducting the special issue of the journal, he expresses the hope that researchers will consider focusing their attention on the security and privacy of IoT.

The security of IoT has received a great deal of attention. A number of reviews have suggested mechanisms to overcome the security threats and challenges of IoT [65], [82], [14], [80], [38]. Most of these reviews have concluded with a set of security practices that should be deployed by IoT product designs. This list usually includes: (1) secure booting using cryptographically generated digital signatures; (2) deploy authentication and access control techniques based on the lightweight public key authentication technology and asymmetric cryptosystems; (3) firewalls; (4) assiduous patching. Finally, they call for increased user awareness of security aspects of IoT [82]. Privacy has received far less attention from researchers.

One systematic review of privacy threats related to IoT was conducted by Ziegeldorf in 2014 [84]. He first classified the evolving technologies used in IoT as: to RFID, wireless sensor network (WSN), smart phones, and cloud computing. He then highlights features that can be considered most important in the context of privacy. These include data collection, life cycle and system interaction. The author studied and analyzed seven threat categories: identification, localization and tracking, profiling, privacy-violating interaction and presentation, life-cycle transitions, inventory attack, and linkage. The study identified privacy-preserving approaches from related work to determine whether they could mitigate in an IoT context.

The author concluded that identification, tracking and profiling were the primary threats that are exacerbated in IoT. The remaining four threats of privacy-violating interactions and presentations, lifecycle transitions, inventory attacks and information linkage are recent additions, prompted by the rise of IoT.

This systematic literature review extends Ziegeldorf's work because his paper focused on analyzing the challenges and threats of IoT in the context of entities and information flows. This paper examines IoT-specific solutions, and identifies gaps in the research literature, specifically from an end-user perspective.

## VIII. CONCLUSION

The era of the Internet of Things has arrived. Current research is disproportionally focused on the security concerns of IoT. Yet the privacy problem is equally urgent. Future research should assess privacy perceptions related to IoT, to find out whether people would act to protect their own privacy when using IoT. Moreover, we should determine whether they would value and use a management tool that explicitly prevents privacy invasions by IoT devices, especially if some degree of effort is involved.

## ACKNOWLEDGEMENTS

A shorter version of this paper will appear in the Proceedings of the Hawaii International Conference on System

Sciences HICSS-50: January 4-7, 2017 — Hilton Waikoloa Village.

TABLE II: Privacy principles addressed by each IoT privacy-preserving solution.

**Privacy Principles**

| Ref | Choice/Consent | Purpose specification | Collection limitation | Data minimisation | Onward transfer | Integrity/Accuracy | Notice/Awareness | Access/Participation | Enforcement/Redress | Security | Use limitation |
|---|---|---|---|---|---|---|---|---|---|---|---|
| **Cryptographic techniques** | | | | | | | | | | | |
| [1] | • | | • | | • | | • | • | • | • | • |
| [2] | • | | • | | • | • | • | | • | • | • |
| [5] | • | | | | | • | • | | | • | |
| [7] | | • | | • | • | | • | | | • | |
| [9] | • | | | | | • | • | • | | • | • |
| [13] | • | | | | • | • | | • | | • | • |
| [18] | • | | | | | | • | | | • | • |
| [20] | • | | | | | | • | | • | • | • |
| [22] | | | | | • | • | | | | • | |
| [24] | • | | | | | • | | • | | • | • |
| [25] | • | • | • | • | • | • | • | • | • | • | • |
| [26] | • | | | | • | • | | • | | • | • |
| [30] | • | • | • | | | • | • | • | • | • | |
| [36] | | • | | • | • | | | | | • | |
| [41] | • | • | • | | • | | • | • | | • | • |
| [43] | | | | | • | | | | | • | |
| [19] | • | | • | • | • | • | • | • | • | • | • |
| [44] | | | • | | | • | | | | • | |
| [45] | • | | | | • | • | | • | | • | • |
| [47] | • | | | | | | • | | | • | • |
| [51] | • | • | | | • | • | • | | | | |
| [53] | | | | | | • | | | | | |
| [54] | • | • | • | | • | • | • | • | • | • | • |
| [56] | | | • | | • | • | • | | • | • | |
| [59] | • | | • | • | | • | • | • | | • | • |
| [58] | • | | | • | • | • | | • | | • | • |
| [64] | • | | | | • | • | | | | • | |
| [67] | | | • | • | • | • | | | | | |
| [77] | | • | • | | • | • | • | | | • | • |
| [73] | | | | | | • | | | | • | |
| [79] | | | • | | | | • | | | • | |
| **Data minimization** | | | | | | | | | | | |
| [7] | | • | | • | • | | • | | | • | |
| [15] | | | | • | | | | | | • | |
| [19] | • | | • | • | • | • | • | • | • | • | • |
| [25] | • | • | • | • | • | • | • | • | • | • | • |
| [59] | • | | • | | • | | • | • | | • | • |
| **Access control** | | | | | | | | | | | |
| [1] | • | | • | | • | | • | • | • | • | • |
| [9] | • | | | | | • | • | • | | • | • |
| [13] | • | | | | • | • | | • | | • | • |
| [19] | • | | • | • | • | • | • | • | • | • | • |

| Ref | Choice/Consent | Purpose specification | Collection limitation | Data minimisation | Onward transfer | Integrity/Accuracy | Notice/Awareness | Access/Participation | Enforcement/Redress | Security | Use limitation |
|---|---|---|---|---|---|---|---|---|---|---|---|
| [20] | • | | | | | • | | • | | • | • |
| [25] | • | • | • | • | • | • | • | • | • | • | • |
| [26] | • | | | | • | • | | • | | • | • |
| [29] | • | | • | | | | | • | | • | • |
| [30] | • | • | • | | | • | • | • | • | • | |
| [31] | • | • | • | | • | | • | • | • | • | • |
| [39] | • | | • | | | • | • | • | | • | • |
| [41] | • | • | • | | • | | • | • | | • | |
| [45] | • | | | | • | • | | • | | • | • |
| [47] | • | | | | | | • | | | • | • |
| [59] | • | | • | • | | • | • | • | | • | • |
| [54] | • | • | • | | • | • | • | • | • | • | • |
| [58] | • | | | • | • | • | | • | | • | • |
| [57] | • | | • | | • | • | | • | | • | |
| [72] | • | • | • | | • | • | • | • | • | • | |
| [74] | • | | • | | • | • | | | | • | |
| **Differential Privacy** | | | | | | | | | | | |
| [19] | • | | • | • | • | • | • | • | • | • | • |
| **Privacy awareness** | | | | | | | | | | | |
| [1] | • | | • | | • | | • | • | • | • | • |
| [2] | • | | • | | • | • | • | | • | • | • |
| [4] | | | | | | • | | | | • | |
| [5] | • | | | | • | • | | | | • | |
| [7] | | • | | • | • | | • | | | • | |
| [8] | | • | • | | | • | | • | | | • |
| [11] | | | • | | | • | | | | | • |
| [25] | • | • | • | • | • | • | • | • | • | • | • |
| [31] | • | • | • | | • | | • | • | • | • | • |
| [33] | | | | | • | • | | | | • | • |
| [41] | • | • | • | | • | | • | • | | • | |
| [54] | • | • | • | | • | • | • | • | • | • | • |
| [55] | | • | | | • | • | | | | • | |
| [56] | | | • | | • | • | • | | • | • | |
| [59] | • | | • | • | | • | • | • | | • | • |
| [70] | | | • | | | | • | | • | | • |
| [69] | | | • | | | | • | | • | | • |
| [71] | | | • | | • | | • | | • | • | |
| [72] | • | • | • | | • | • | • | • | • | • | |
| [77] | | • | • | | • | • | | | | • | • |
| [81] | | | | | • | • | | | | • | |
| **Other** | | | | | | | | | | | |
| [6] | | | | | • | • | | | | • | |
| [34] | | | | | • | • | | | | • | |
| [40] | | | | | • | • | | | | • | |
| [68] | | | | | • | • | | | | • | |
| [83] | | | | | • | • | | | | • | |

TABLE III
CHARACTERISATION OF IOT PRIVACY-RELATED PAPERS

| | Total | Europe | Other |
|---|---|---|---|
| **Methods Used** | | | |
| Observation | 12 | 4 | 8 |
| Surveys | 15 | 8 | 7 |
| Interviews | 1 | 1 | 0 |
| Focus groups | 0 | 0 | 0 |
| Field research | 0 | 0 | 0 |
| Case studies | 18 | 8 | 10 |
| Document analysis | 20 | 9 | 11 |
| Modelling | 52 | 19 | 33 |
| Unspecified | 29 | 2 | 27 |
| **Type of Data** | | | |
| Qualitative | 24 | 10 | 14 |
| Quantitative | 55 | 22 | 33 |
| Mixed | 20 | 8 | 12 |
| News or reports | 22 | 1 | 21 |
| **Application Areas for the Internet of Things** | | | |
| Home automation | 47 | 11 | 36 |
| Smart cities | 21 | 11 | 10 |
| Smart manufacturing | 12 | 6 | 6 |
| Health care | 16 | 5 | 11 |
| Automotive | 17 | 6 | 11 |
| Wearables | 12 | 4 | 8 |
| Unspecified | 62 | 23 | 39 |
| **Technologies of IoT** | | | |
| RFID | 36 | 14 | 22 |
| Sensor technology | 57 | 18 | 39 |
| Nano technology | 1 | 0 | 1 |
| Intelligence embedded technology | 9 | 6 | 3 |
| Unspecified | 51 | 17 | 34 |
| **Privacy Protection** | | | |
| Cryptographic techniques and information manipulation | 62 | 23 | 39 |
| Data minimization | 13 | 6 | 7 |
| Access control | 40 | 17 | 23 |
| Privacy awareness or context awareness | 41 | 16 | 25 |
| Differential Privacy | 1 | 1 | 0 |
| Other | 16 | 6 | 20 |
| **Privacy threats** | | | |
| Identification | 28 | 9 | 19 |
| Location and Tracking | 34 | 9 | 25 |
| Profiling | 23 | 9 | 14 |
| Interaction & Presentation | 7 | 2 | 5 |
| Lifecycle transitions | 4 | 3 | 1 |
| Inventory attack | 9 | 3 | 6 |
| Linkage | 3 | 1 | 2 |
| Unspecified | 71 | 26 | 45 |
| **Privacy Perceptions** | | | |
| Fundamentalist (most concerned) | 112 | 39 | 73 |
| Pragmatic (less concerned) | 6 | 2 | 4 |
| Unconcerned (least concern) | 2 | 1 | 1 |
| Unspecified | 6 | 2 | 4 |
| **Privacy or Security Violations** | | | |
| Accidental or inadvertent violation | 1 | 0 | 1 |
| Failure to follow established privacy and security policies and procedures | 1 | 0 | 1 |
| Deliberate or purposeful violation without harmful intent | 15 | 2 | 13 |
| Willful and malicious violation with harmful intent | 26 | 6 | 20 |
| Unspecified | 82 | 33 | 49 |